\documentstyle[preprint,amsfonts,amsmath,prd,aps,12pt,epsf,epsfig]{revtex}
\begin{document}
\preprint{
\vbox{
\halign{&##\hfil\cr
         & HUPD-0166\cr
         & hep-ph/0104309\cr
         & April 2001 \cr
         & \cr
         & \cr
}}}
\vskip 4mm
\title{Double $J/\psi$ Production at Photon Colliders}
\author{Cong-Feng Qiao\footnote{JSPS Research Fellow. E-mail:
qiao@theo.phys.sci.hiroshima-u.ac.jp}}
\vskip 1pt
\address{Department of Physics, Faculty of Science,\\
Hiroshima University, Higashi-Hiroshima 724, Japan}
\maketitle
\vskip 2.5cm
\centerline{\bf Abstract}
\vskip 7mm
\begin{minipage}{145mm}
The double $J/\psi$(DJ) production in direct photon-photon collision 
is investigated. It is found that the $J/\psi$ production rate in 
this process is of the same order of magnitude as those of 
previously discussed ones, which hints the dominant $J/\psi$ 
inclusive production process in the direct photon-photon
collision is still not being touched. As DJ production 
process has a clear distinction from the single $J/\psi$ 
inclusive processes in the final states, 
experimental study on it would be helpful to clarify the 
validity of color-singlet model.
\vskip 9pt

\noindent
PACS Number(s){12.38.Bx, 13.65.-i, 14.40.Lb}
\end{minipage}
\vfill \eject
The quarkonium production and decays have long been taken as an ideal 
means in investigating the nature of QCD and other new phenomena. 
Hence, to establish a proper theory which can precisely describe 
the heavy quarkonium production and decays is very necessary. 
The novel effective theory, the non-relativistic QCD(NRQCD) 
\cite{nrqcd}, is possibly the one to this aim, which is formulated 
from the first principles. To make a precise prediction 
for quarknoium production, however, at now, with only the 
NRQCD is not enough since the magnitude of the non-perturbative 
parameters in the theory are still unknown. Or in other words, 
results are always sensitive to the values of the input 
non-perturbative parameters, whereas the NRQCD can at most give 
the relative weights of these parameters in order of $v^2$ based 
on the "velocity scaling rules". The present situation 
in the Onium physics is that on one side the color-octet model 
\cite{com} stands as the most plausible approach, till now, in 
explaining the large transverse momentum $\psi(\psi')$ production 
"anomaly"; on the other hand, the model meets some difficulties in 
confronting with other phenomena \cite{rothstein}. Therefore, to what
degree the color-octet mechanism plays the role in quarkonium 
production is still not clear and interesting. Either to find 
distinctive signatures of Color-octet or more strong arguments 
against it is currently an urgent task in this research realm. 
To this end lots of works have been done, however, unfortunately 
up to now there is still no definite answer.

In recent years several new concepts on linear colliders aiming at
providing collisions at the center-of-mass energy from hundreds GeV 
to multi-TeV with high luminosity are proposed and the feasibilities 
are pre-tested, such as JLC at KEK, TESLA at DESY and CLIC at CERN, 
etc. Theoretically, high-intensity photon beams may be obtained 
by the Compton back-scattering of laser light off the linac electron 
beams and realize the photon-photon collision with approximately the 
same luminosity as that of the $e^+$ $e^-$ beams. Such a photon linear
collider can have high energy up to the TeV order. Since the first
discovery of $J/\psi$ in 1974, researches on quarknoium production 
at $e^+$ $e^-$ colliders at various energies have been carried out 
in details \cite{ee1,ee2,ee3,ee4,ee5}. However, studies on the 
photon-photon scattering are very limited and just begin 
\cite{gg1,gg2,gg3}. One of the difficulties one may meet in doing
a complete calculation for two photon process beyond LO in $\alpha$,
the electromagnetic coupling constant, or $\alpha_s$, the strong 
coupling constant, is how to simplify the lengthy expressions in order
to present them analytically. In this brief report we study the 
double quarkonium, the $J/\psi$, production in direct photon-photon
collision(of the resolved case see, e.g., ref. \cite{ee5}) at the 
order $\alpha^2\alpha_s^2$. From physical point of view, in this
process one may get rid of the inaccuracy induced by the uncertainties
of the color-octet matrix elements 
and by the contributions from higher excited 
states as well, because they are much suppressed relative to the 
direct contribution in color-singlet.
 
In $\gamma\gamma$ scattering, at leading order in $\alpha$ the $J/\psi$
is produced via the process
\begin{eqnarray}
  \label{eq:leading1}
  \gamma + \gamma \rightarrow J/\psi + \gamma\;.
\end{eqnarray}
However, since at the scale of heavy quark mass, the strong coupling 
constant is not too small, the process 
\begin{eqnarray}
  \label{eq:leading2}
  \gamma + \gamma \rightarrow J/\psi^{(8)} + g
\end{eqnarray}
may compete with the pure electromagnetic process (\ref{eq:leading1}) 
through the Color-Octet mechanism \cite{gg1}. Here, we schematically 
use the $J/\psi^{(8)}$ to denote those evolved from the Color-Octet 
states. 

To go one order up in $\alpha_s$, one may still expect to get the 
same order of maginitude in $J/\psi$ production rate, because in this 
case the $J/\psi$ may be produced in color-singlet and therefore will 
get compensation for the $\alpha_s$ suppression from the 
non-perturbative sector relative to the octet process 
(\ref{eq:leading2}). What we proceed here, the DJ production in 
$\gamma\gamma$ collision, is a sub-category of the inclusive 
$J/\psi$ production process at order $\alpha^2\alpha^2_s$. That is 
\begin{eqnarray}
  \label{eq:leading3}
  \gamma(k_1) + \gamma(k_2) \rightarrow J/\psi(P) + J/\psi(P')\;, 
\end{eqnarray}
as shown in Figure 1, which consists of twenty Feynman diagrams. 
Here, the momenta of the related particles are shown in the 
parentheses explicitly. Similar as the situation of $B_c$ production 
in photon-photon collision \cite{klr}, these twenty diagrams 
involved here can also be classified as two sub-groups, (A) and (B), 
and each of them are gauge invariant. 

The differential cross-section of process (\ref{eq:leading3}) is 
obtained as
\begin{eqnarray}
\label{diff}
\frac{d \sigma}{d t} &=& \frac{262144\alpha^2\alpha_s^2 \pi |R(0)|^4}
{729 m^2 s^6 (m^2 - t)^4  (m^2 - u)^4}
\left[296  m^{20} - 1112  m^{18} (t + u) + 
 t^4  u^4 (t + u)^2 - 8  m^2  t^3 u^3  (t + u)^3 \right.\nonumber \\
& + &  m^{16}  (1917  t^2 + 3578  t  u + 1917  u^2) - 
8  m^{14}  (242  t^3 + 657  t^2  u + 657  t u^2 + 242  u^3)\nonumber \\
& + & 2  m^{12}  (623  t^4 + 2204 t^3  u + 3298  t^2  u^2 + 
2204  t  u^3 + 623 u^4) \nonumber \\ 
& - & 4  m^{10}  (131  t^5 + 563  t^4  u + 1156  t^3  u^2 + 
1156  t^2  u^3 + 563  t  u^4 + 131  u^5) \nonumber \\ 
& + &
m^8  (143  t^6 + 690  t^5  u + 1899 t^4  u^2 + 2640  t^3  u^3 + 
1899  t^2  u^4 + 690  t  u^5 + 143  u^6) \nonumber \\
& - & 
4  m^6  (6  t^7 + 29  t^6  u + 107  t^5  u^2 + 212  t^4  u^3 + 
212  t^3  u^4 + 107  t^2  u^5 + 29  t  u^6 + 6 u^7) \nonumber \\
& + & 
\left. 2  m^4  (t^8 + 4  t^7  u + 21  t^6 u^2 + 68  t^5  u^3 + 
104 t^4 u^4 + 68  t^3  u^5 + 21  t^2  u^6 + 4  t  u^7 + u^8)\right]\;.
\end{eqnarray}
In obtaining the above analytical expression, we start from the general
Feynman rules and project the Charm-anti-Charm pair into the S-wave 
vector Onium state in color-singlet. To manipulate the trace and matrix 
element square of those twenty diagrams, the computer algebra system 
MATHEMATICA is used with the help of the package FEYNCALC 
\cite{feyncalc}. 

The total cross section of the direct photon-photon collision can be 
obtained by convoluting the differential cross-section with the photon 
distribution functions, like
\begin{eqnarray}
\sigma_{total} = \int d t dx_1 dx_2 f_\gamma(x_1) f_\gamma(x_2) 
\frac{d\sigma_{\gamma\gamma}}{d t}(x_1, x_2)\;,
\end{eqnarray}
where $\frac{d\sigma_{\gamma\gamma}}{d t}(x_1, x_2)$ is the 
differential cross-section given in (\ref{diff}); the 
$f_\gamma(x_i)$, $i = 1, 2$, is the photon distribution with 
the fraction $x_i$ of the beam energy,
\begin{eqnarray}
f_\gamma(x) =  \frac{1}{N}\left[1 - x + \frac{1}{1 - x} - 4 r 
(1 - r)\right]\;.
\end{eqnarray}
Here, $r \equiv x/(x_m (1 - x))$, the normalization
\begin{eqnarray}
N =  (1 - \frac{4}{x_m} - \frac{8}{x_m^2})\log(1 + x_m) + 
\frac{1}{2} + \frac{8}{x_m} - \frac{1}{2 (1 + m_x)^2}
\end{eqnarray}
and
\begin{eqnarray}
x_m =  \frac{4E_b E_l}{m_e^2}\cos^2\frac{\theta}{2} ~,
\end{eqnarray}
where $E_b$ and $E_l$ are the energies of electron beam and 
laser photon, respectively, and the $\theta$ is the angle between them. 
The energy fraction $x$ of the photon is restricted in
\begin{eqnarray}
0 \le x \le \frac{x_m}{1 + x_m}~,
\end{eqnarray}
and the maximum energy of the obtained photon beam depends on $x_m$, 
i.e.,
\begin{eqnarray}
\omega_{max} = \frac{x_m}{1 + x_m} E_b~.
\end{eqnarray}
In our numerical calculation we take the optimum value of 
$x_m = 4.83$ in avoiding the background $e^+ e^-$ pair production 
from the laser and backscattered photon collision \cite{telnov}. 

Other values of input parameters used in our numerical calculation are:
\begin{eqnarray}
m_c =  1.5\; \rm{GeV},\; \alpha = 1/128,\; \alpha_s(m_c) = 0.3,\; 
|R(0)|^2 = 0.8\; \rm{GeV}^3\;, <{\cal{O}}^{J/\psi}_8> = 0.01 
\rm{GeV}^3\;.
\end{eqnarray}
With the above formulas and input parameters one can immediately get
the total cross-sections. For instance, the $J/\psi$ production 
cross-sections of the processes (\ref{eq:leading1}), (\ref{eq:leading2}) 
and (\ref{eq:leading3}) are 2.54 fb, 1.92 fb and 2.15 fb, respectively 
with colliding energy being of 500 GeV. Since the projected linear 
colliders with luminosity of hundreds fb$^{-1}$ per year, for example 
the TESLA, and the integrated total cross-sections increase with the 
colliding energy decreasing as shown in Figure 2, we may have 
hundreds of events being observed in one year at colliding energy 
500 GeV or less. However, when the colliding energy goes up to 1TeV 
the cross-sections for processes (1) -- (3) would be three orders 
less than those of them at 500 GeV, respectively, hence, all of 
them would be unobservable. In Figure \ref{graph2} the energy 
dependences of the total cross-sections are shown from about 
the energy of LEP II to the next generation of linear collider 
energy. It can be seen that the color-octet process is the 
smallest one all over this energy scope in the concerned three 
processes, though they are in the same order. 

Another thing we want to point out here is that the DJ 
production discussed in this work can not be simply explained 
by the quark fragmentation. With more details: we have the 
differential cross-section of $\gamma\gamma \rightarrow Q\bar{Q}$ as
\begin{eqnarray}
\frac{d \sigma}{d t} &=& \frac{- 2 \pi \alpha^2 e_Q^4}
{s^2 (m_Q^2 - t)^2 (m_Q^2 - u)^2}
\left[6  m_Q^8 - t u (t^2 + u^2) - m_Q^4 (3 t^2 + 14 t u + 3 u^2) 
\right. \nonumber\\
&+&
\left. m_Q^2 (t^3 + 7 t^2 u + 7 t u^2 + u^3) \right]\;,
\end{eqnarray}
and know the probability of Charm quark evolving to $J/\psi$ is of 
$1.6 \times 10^{-4}$ via fragmentation \cite{bcy}. To a certain
degree of approximation, one can get the DJ 
production rate through fragmentation mechanism by simply 
combining the Charm production rate with
the universal fragmentation probability. Therefore, we have a
cross-section of $1.1 \times 10^{-3}$fb at energy 500 GeV, 
which is much smaller than that of the complete calculation of 
the DJ process. The reason of this is that the fragmentation 
mechanism only applicable when the fragmenting quark possesses 
large momentum and small off-shellness. In the double Onium production 
this condition can not be fulfilled for both fragmenting quark at 
the same time. On the other hand, one can know from the above 
discussion that the cross-section for single $J/\psi$ production 
via fragmentation mechanism would be about one order larger than 
those of the processes (1) -- (3) at 500 GeV. Furthermore, we get 
know that the process at order $\alpha^2\alpha_s^2$ for single 
$J/\psi$ inclusive production should be the dominant one in 
photon-photon collision, which seems to be overlooked.

Considering there are remaining disagreements among three independent 
calculations on the leading order processes (\ref{eq:leading1}) and 
(\ref{eq:leading2}) \cite{gg1,gg2,gg3}, we have done a check over 
them and find an agreement with refs.\cite{gg1} and \cite{gg3} on 
the expression for octet process, i.e., the eq.(7) of ref.\cite{gg1}. 
And our replacement for obtaining the singlet process agrees with 
that in ref \cite{gg1}.

In conclusion, we have calculated the DJ 
production at photon colliders. It is found that at moderate 
energy of the next generation linear colliders there would be 
hundreds of events to be detected per year with the high projected 
luminosity. In addition, since the production rates of $J/\psi$
via the color-octet mechanism, electromagnetic process, and what 
discussed here are almost the same in the full scope of the 
colliding energy, to differentiate the color-octet mechanism from 
the color-singlet one in these processes, experimentally one 
should detect not only the $J/\psi$ but also other final states. 
Relatively the processes (1) and (3) are more liable to be discerned 
than (2), because the signal of the latter is easily to be entangled 
with that of the single $J/\psi$ production process at 
the order $\alpha^2\alpha^2_s$ \cite{cfq}. 

Like in the case of double $\pi$ production at photon colliders, 
where the Pion wavefuctions and the photon-to-pion transition 
form factor may be detected \cite{sjb}, the DJ process 
may provide more information on the color-singlet description 
of quarkonium production considering that it gets less
influences from the color-octet uncertainties and the higher 
excited states feeddown.

In the end, with the expression (4) similar process in the soft 
bremsstrahlung photon collision can be easily discussed with 
the Weiz\"acker-Williams approximation. 

\vskip 1.2cm
\centerline{\bf ACKNOWLEDGEMENTS}
\vskip 0.3cm
This work is supported by the Grant-in-Aid aid of JSPS committee and 
the author would like to express his gratitude to NSF of China for 
support, ITP for hospitality, and J.P. Ma for kind invitation for the 
visit, while this work initiated, and also to Stanley J. Brodsky for 
helpful comments and discussions.

\begin{figure}
\vskip -3cm
\epsfxsize=15 cm
\centerline{\epsffile{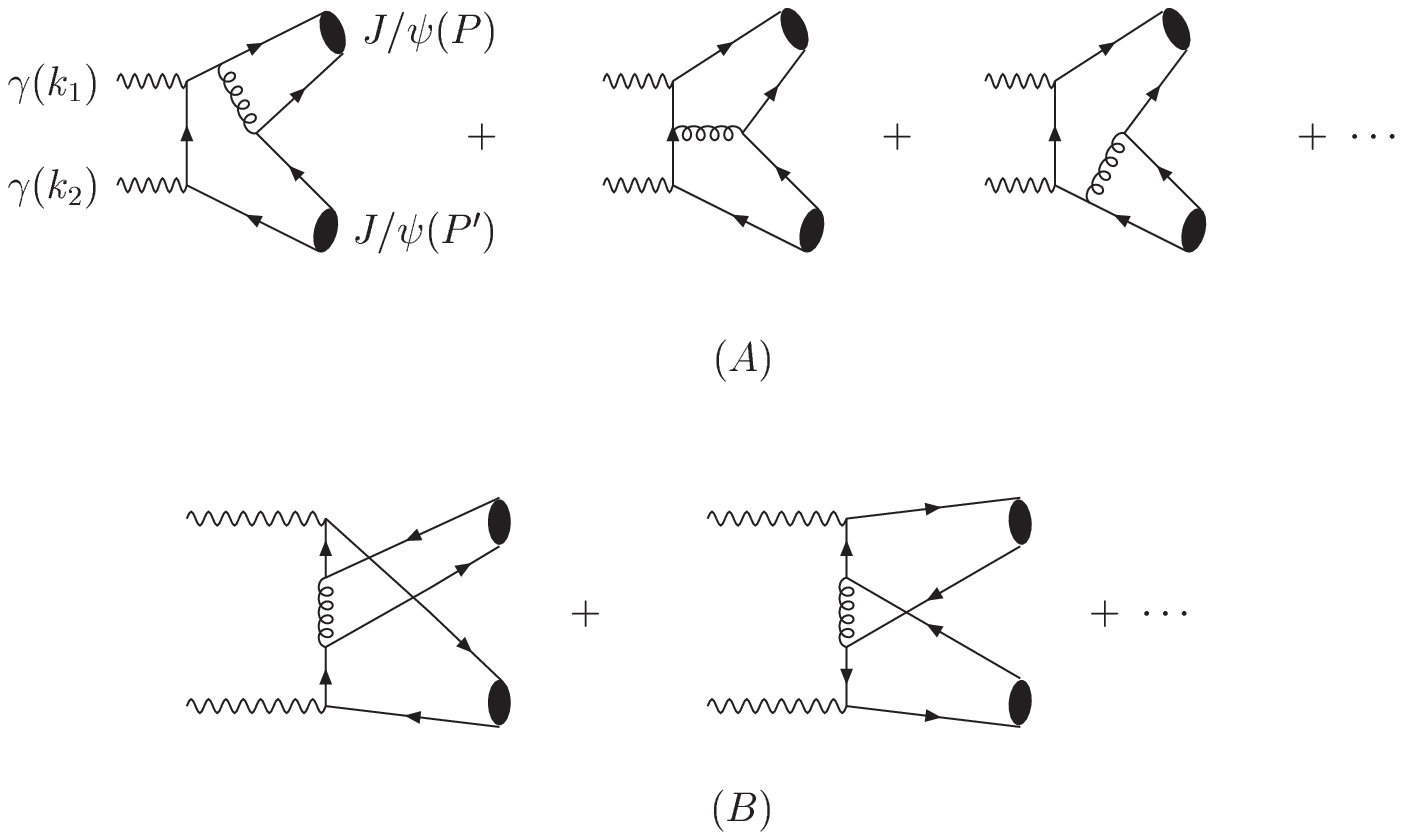}}
\vskip -10cm
 \caption[]{The Feynman diagrams of the double $J/\psi$ production in
$\gamma\gamma$ collision.}
\label{graph1}
\end{figure}    

\begin{figure}[tbh]
\begin{center}
\epsfig{file=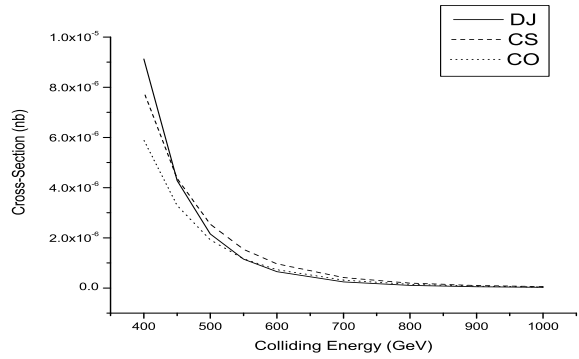,bbllx=90pt,bblly=300pt,bburx=230pt, 
bbury=420pt,width=4cm,height=2.5cm,clip=0}
  \end{center}
\vskip 6cm
  \caption[bt]{The engergy dependence of the cross-sections. DJ: the 
 double $J/\psi$ process; CS: the color-singlet process 
 (\ref{eq:leading1}); CO: the color-octet process (\ref{eq:leading2}).}
  \label{graph2}
\end{figure}    
\end{document}